\documentclass[prc,amsmath,twocolumn,showpacs,superscriptaddress]{revtex4}
\usepackage{graphicx}
\usepackage{amssymb}
\usepackage{enumerate}
\usepackage{verbatim}

\begin{document}

\title{Finite range effects in (d,p) reactions}

\author{N.~B.~Nguyen}
 \affiliation{National Superconducting Cyclotron Laboratory, Michigan State University, East Lansing, Michigan 48824}
 \affiliation{Department of Physics and Astronomy, Michigan State University, East Lansing, MI 48824-1321}
\author{F.~M.~Nunes}
 \affiliation{National Superconducting Cyclotron Laboratory, Michigan State University, East Lansing, Michigan 48824}
 \affiliation{Department of Physics and Astronomy, Michigan State University, East Lansing, MI 48824-1321}
\author{R.~C.~Johnson}
 \affiliation{Department of Physics, University of Surrey, Guildford GU2 7XH, U.K.}
 \affiliation{National Superconducting Cyclotron Laboratory, Michigan State University, East Lansing, Michigan 48824}
 \affiliation{Department of Physics and Astronomy, Michigan State University, East Lansing, MI 48824-1321}

\date{\today}

\begin{abstract}
With the increasing interest in using (d,p) transfer reactions
to extract structure and astrophysical information, it is important to
evaluate the accuracy of common approximations in reaction
theory. Starting from the zero-range
adiabatic wave model, which takes into account deuteron breakup in the
transfer process, we evaluate the importance of the finite range of the $n-p$ interaction
in calculating the adiabatic deuteron wave (as in Johnson and Tandy) as well as in evaluating
the transfer amplitude. Our study covers a wide variety of targets
as well as a large range of beam energies.
Whereas at low beam energies finite range effects
are small (below 10\%), we find these effects to become important at
intermediate energies (20 MeV/u) calling for an exact treatment of finite range
in the analysis of (d,p) reactions measured at fragmentation facilities.
\end{abstract}



\maketitle
\section{Introduction}

Since the early days of nuclear physics, transfer reactions have been
a preferred tool to study spectroscopic properties of nuclei and
have been widely used to determine single particle
structure across the nuclear chart (e.g. \cite{niewo66,daehnick69,bainum77,schiffer04}).
Such studies have allowed a better understanding of detailed features of 
nuclear interactions. It was through a systematic study of single nucleon transfer
on Sn isotopes, that we now understand the reduction of the spin-orbit strength
when moving toward neutron rich systems \cite{schiffer04}.
One would also like to use the transfer reaction method to discriminate between
effective interactions used in the shell model, as suggested in the study
of (d,p) on Ni isotopes \cite{lee09}.
Nowadays the spectroscopy of exotic nuclei can be studied
using inverse kinematics. Pioneering studies \cite{li8dp,he6dp,li9dp,ge83se85,Catford2005}
hold promise for applying this technique more broadly, especially in new
generation rare isotope facilities, where beam rates will be enhanced.

Another intriguing aspect of nuclear structure is the role of correlations.
The independent particle shell model of course neglects all correlations.
State-of-the-art shell model include some correlations effectively through the
residual interactions.
Electron knockout experiments have shown a 30\% decrease in the spectroscopic
factors of closed shell nuclei as compared to the independent shell model predictions,
a reduction that is understood in terms of short range correlations \cite{kramer}.
Reduction factors from nuclear knockout experiments can be much larger,
depending on the difference between the neutron and proton separation energies
\cite{lee09prl}.
However, the physical reason for such a large suppression is still unclear,
long range correlations being a possibility \cite{barbieri}.
What is most intriguing is that spectroscopic factors from transfer reactions
show no dependence on the difference in neutron-proton separation energies.
Large surveys of ground state spectroscopic factors from (d,p) reactions \cite{tsang05},
including nuclei with a wide range of separation energies (0.5-19 MeV) show
good agreement with large scale shell model using modern effective interactions.

The reconciliation of these results with those from
knockout experiments\cite{gade08} is proving difficult and is the subject of recent work on new approaches to the calculation of overlap functions\cite{Timofeyuk2009,barbieri}.
The resolution of these important physics questions relies on the accuracy of the reaction model
being used. It is thus of paramount importance that the reaction theory is well
founded and uncertainties are well understood. In this respect there have been
a number of works looking at specific aspects of the reactions theory used for the
analysis of (d,p) transfer data (uncertainties in the optical potentials \cite{liu04},
coupling to excited states of the target \cite{delaunay05}
and ambiguities due to the single particle wave functions \cite{muk05,pang06} and new ways of calculating overlap functions \cite{Timofeyuk2009,barbieri}). In this work we explore another aspect: the consequences of the non-zero range of the $n-p$ interaction that plays a key role in the theory is several different ways.

Over the past four decades of work on (d,p) reactions, different approximations were made,
not only regarding the optical potentials, the deuteron and final single-particle wavefunctions,
but also in the reaction mechanism and the
evaluation of the transfer matrix element. It may thus be confusing to realize
the large range of spectroscopic factors available in the literature.
Systematic and consistent studies  \cite{tsang05,lee07,tsang09,lee09} provide a much
better overall assessment of the situation. In \cite{tsang05,lee07} an extensive survey of
ground state spectroscopic factors from charge $Z=3-24$ was performed using the same
reaction model, the same global potentials and the same single particle parameters.
With these same assumptions,
spectroscopic factors extracted from 235 sets of (d,p) data were found consistent
with shell model predictions to within $\pm 20$\%.
An identical study was performed on excited states \cite{tsang09}, and although there
were a few unresolved cases (such as the Ni isotopes),
the overall agreement with shell model was of the order of $40$\%.
Ni isotopes were studied separately \cite{lee09} and a better overall description
of these nuclei was found with a modified effective interaction in the shell-model calculations.

All these studies
rely on the adiabatic distorted-wave approximation  (ADWA) developed by Johnson and Soper \cite{soper}
and the local energy approximation (LEA) \cite{lea} to take into account the finite range of the $n-p$ interaction $V_{np}$ in evaluating the $(d,p)$ transition matrix.

In \cite{soper}, a three-body theory for
A(d,p)B was developed taking into account deuteron breakup which is known to be
important even for reactions on stable nuclei.
In \cite{soper} the zero range approximation is made for the $n-p$ interaction.
Then, the transfer matrix element reduces to a form similar to the zero range Distorted
Wave Born Approximation (DWBA) where the  scattering wavefunction in the incident channel is calculated
with an adiabatic potential consisting of the sum of the proton and neutron potentials
evaluated at half the deuteron energy, instead of the deuteron optical potential extracted
from (d,d) data.
ADWA typically decreases the radius and diffuseness of
the distorting potential \cite{harvey} compared with typical values for deuteron optical potentials deduced from elastic deuteron scattering  and often
provides a better description of the data \cite{johnson-ria}.

An adiabatic theory including finite range effects was formally developed by Johnson and Tandy \cite{tandy}.
Using a Weinberg expansion in the deuteron channel, Johnson and Tandy arrive at a set of coupled
channel equations to describe the relative motion of the centre of mass of the neutron and proton relative to the target. The solution of the coupled channels equations gives a three-body wavefunction that is a coherent superposition of the bound (deuteron) and break-up continuum states of the $n-p$ system.

In this paper we will confine ourselves to the simplest version of this theory in which only the first term in the Weinberg expansion is retained. This assumes that the only significant break-up components in the three-body wavefunction have sufficiently small energies that \emph {inside the range of} $V_{np}$, the relevant $n-p$ scattering states have the same radial shape as the deuteron ground state wave function. 
When only the first Weinberg term is included, the coupled equations reduce to an optical-model-like equation for the three-body wavefunction
where the distorting potential is the sum of the neutron and proton optical
potentials multiplied by the neutron-proton interaction, folded over the deuteron
bound state.The effect of other components \cite{laid} have been shown to be significant at $E_d=88$ MeV but their effects at lower energies are unknown.

In the early sixties, it was already understood that finite range effects
were important in (d,p) reactions, however due to computational limitations,
all calculations were performed in zero range. A very popular procedure to correct a zero
range calculation was developed by Buttle and Goldfarb \cite{lea}.
The standard implementation of this method, the so-called local energy approximation (LEA),
takes only the first term of the expansion presented in \cite{lea}.
For deuteron energies well above the Coulomb barrier,
it is not clear that this procedure is sufficiently accurate.
A simple estimate of the modification of the Johnson and Soper
potential due to a finite range $V_{np}$ were reported in \cite{wales}.

In this work, we perform a systematic study of finite range
effects in (d,p) reactions within the framework of ADWA.
We consider 26 (d,p) reactions on stable targets,
involving nuclei with masses ranging $A=12-208$ and deuteron energies $E_d =2-70$ MeV.
We first study the finite range effects on the  distorting
potential potential in the incident channel following the method by Johnson and Tandy \cite{tandy}. In addition,
we consider  finite range effects in the evaluation of the transfer matrix element and the accuracy of the LEA.
We also explore the implications of our study to reactions
involving loosely bound nuclei.

The paper is organized in the following way: reaction theory is revised in Sec. II,
results are presented in Sec. III and further discussed in section IV.
Finally, conclusions are drawn in Sec. V.

\section{Theory}

Our starting point is a three-body model of the $n+p+A$ system, see \cite{book}, Ch.7 . In this model the scattering wavefunction $\Psi^{(+)}(\vec r,\vec R)$ corresponding to a deuteron incident on a nucleus $A$ is the solution of the inhomogeneous differential equation
\begin{eqnarray}
[E+i\epsilon - T_\mathbf{r}- T_\mathbf{R}  - U_{nA} &-& U_{pA} - V_{np} ] \Psi^{(+)}(\vec r,\vec R)
\nonumber \\
&=&\imath \epsilon \phi_{d}(\vec r)\exp(\imath \vec K_d .\vec R).
\label{3b-eq}
\end{eqnarray}
Here $T_\mathbf{r}$ and $T_\mathbf{R}$ are the total kinetic energy operators
associated with the n-p relative motion
and the  motion of the $n+p$ centre of mass relative to the target, where $\vec r = \vec r_p - \vec r_n$ and $\vec R = (\vec r_n + \vec r_p)/2$.
We take $\vec r_n$ and $\vec r_p$ to be the neutron and proton coordinates relative
to the center of mass of the target $A$.
 
In Eq.(\ref{3b-eq}) the interactions $U_{nA}(\vec r_n)$, $U_{pA}(\vec r_p)$ and $V_{np}(\vec r)$
are the neutron-target, proton-target and neutron-proton interactions, respectively.
The term proportional to $\imath \epsilon$ on the r.h.s. of Eq.(\ref{3b-eq}) ensures that there is an incoming wave only in the deuteron channel.

One way of calculating  the  exact (d,p) transition amplitude, $T$, from $\Psi^{(+)}(\vec r,\vec R)$ is to use the formulation of Timofeyuk and Johnson\cite{timofeyuk99}. In the limit $\epsilon \rightarrow 0 $, the transition amplitude  $T$ is given by:
\begin{equation}
T = <\phi_{nA}\tilde{\chi}^{(-)}_{pB}| V_{np} | \Psi^{(+)}>\, ,
\label{tjt-eq}
\end{equation}
where $\phi_{nA}$ is the final state of the neutron-target system, and  $\tilde{\chi}^{(-)}_{pB}$ has a plane wave proton and an incoming scattered wave distorted by $U_{pA}^*$ \cite{timofeyuk99,johnson09}.
Note that in this formulation there is no remnant term and the proton distorted wave is generated by
$U_{pA}^*$ {\em not} $U_{pB}^*$. Eq.(\ref{texact-eq}) neglects recoil effects of order $1/A$ which were evaluated in \cite{timofeyuk99}. They are negligible in all the cases discussed here except possibly for $^{12}$C. The connection between this formulation and standard 3-body methods based on the Faddeev equations \cite{altetal} is explained in\cite{johnson09}.

A more common expression for the transition amplitude for this process is:
\begin{equation}
T = <\phi_{nA}\chi^{(-)}_{pB}| V_{np} + U_{pA} - U_{pB} | \Psi^{(+)}>\, ,
\label{t-eq}
\end{equation}
where $\phi_{nA}$ is the final state of the neutron-target system, and  $\chi_{pB}^{(-)}$ is a
 proton scattering wave distorted by $U_{pB}^*$. In the many-body generalisation of Eq.(\ref{t-eq}) the remnant term $U_{pA} - U_{pB}$ is a function of the internal coordinates of $A$ and $B$ and makes the interpretation of the transition amplitude in terms of nuclear structure (overlap functions) much more complicated than when Eq.(\ref{tjt-eq}) is used. This introduces into the formulation an additional optical potential $U_{pB}$ and thus larger uncertainties into the analysis since in most applications to exotic nuclei this potential is not well determined. Many of the recent applications have used 
Eq.(\ref{t-eq}) and neglected the remnant term $U_{pA} - U_{pB}$:
\begin{equation}
T = <\phi_{nA}\chi^{(-)}_{pB}| V_{np}  | \Psi^{(+)}>\, .
\label{texact-eq}
\end{equation}
Neglecting the remnant term is a very good approximation for all cases discussed here with the exception 
of $^{12}$C. Eq.(\ref{texact-eq}) is the starting point for the present study.

The important realization in \cite{soper,tandy}
is that with Eq.(\ref{tjt-eq}) or Eq.(\ref{texact-eq}), the full three-body wavefunction 
$\Psi^{(+)}(\vec r,\vec R)$ is only required within the range of the $V_{np}$ interaction. 
A major simplification is achieved in the limit of the zero-range approximation as then only 
$\Psi^{(+)}(0,\vec R)$ is needed.

\subsection{Johnson and Soper method}

The method developed by Johnson and Soper \cite{soper} is based on an expansion of
the three-body wavefunction $\Psi^{(+)}(\vec r,\vec R)$ in the complete set of
eigenstates of the $n-p$ Hamiltonian:
\begin{eqnarray}
(T_r + V_{np}) \phi_d(\vec{r}) &=& -\varepsilon_d \phi_d(\vec{r}) \, , \nonumber \\
(T_r + V_{np}) \phi_k(\vec{r}) &=& +\varepsilon_k \phi_k(\vec{r}) \, .
\end{eqnarray}
Here $\phi_d(\vec{r})$ is the deuteron wave function while scattering states
are represented by $\phi_k(\vec{r})$. The three-body wavefunction is
then expanded as:
\begin{equation}
\Psi^{(+)}(\vec r,\vec R)=\phi_d(\vec{r})\chi_d(\vec{R})+\int \mathrm{d}\vec{k}\phi^{(+)}_k(\vec{r}) \chi_k(\vec{R}) \;.
\label{phid-eq}
\end{equation}
When this expansion is introduced in Eq.(\ref{3b-eq}),
and assuming that the excitation energies of the deuteron are small
compared to the deuteron initial kinetic energy $\varepsilon_k +\varepsilon_d << E$,
the three-body equation for $r=0$ reduces to an optical model equation:
\begin{equation}
(E + \epsilon_d - T_R - U_{JS}(R)) \chi^{JS}_d(\vec{R})=0 \, ,
\label{js-eq}
\end{equation}
where the effective potential $U_{JS}$ does not describe deuteron elastic
scattering, but rather incorporates deuteron breakup effects within the
range of $V_{np}$:
\begin{equation}
U_{JS}(R)=U_{nA}(R)+U_{pA}(R) \,.
\label{jspot-eq}
\end{equation}
Within this model, the transfer amplitude reduces to
\begin{equation}
T_{JS}=D_0 \int \mathrm{d}R \, \phi_{nA}^*(\vec{R})\chi_{pB}^*(\vec{R}){\chi}^{JS}_d(\vec{R})
\label{tmatrix-zr}
\end{equation}
where $D_0$ is the zero range constant of the deuteron.

\subsection{The Johnson and Tandy generalisation of the Johnson Soper method}

The Johnson and Tandy \cite{tandy} approach again builds on the fact that
the three-body wavefunction is only needed within the range of $V_{np}$.
While the continuum discretized coupled channel method (CDCC) \cite{cdcc} uses a basis of  eigenstates of the $n-p$ Hamiltonian which is complete for all values of the $n-p$ separation $\vec{r}$, as in Eq.(\ref{phid-eq}),
here the Weinberg basis is introduced:
\begin{equation}
(T_r + \alpha_i V_{np}) \phi_i(\vec{r}) = -\varepsilon_i \phi_i(\vec{r})
\label{weinberg1}
\end{equation}
with $i=1,2,...$ and the orthonormality relation
$\langle \phi_i |  V_{np} | \phi_j \rangle = - \delta_{ij}$.
The Weinberg states (or Sturmians) form a complete basis within the range
of the $V_{np}$ interaction and thus they are particularly suited to
describing the problem when using the transfer amplitude written as in Eq.(\ref{texact-eq}).
A clear advantage of this basis as compared to Eq.(\ref{phid-eq}) is that it is square integrable.

The three-body wavefunction is then expanded as:
\begin{equation}
\Psi^{(+)}(\vec r,\vec R) = \sum_{i=1}^{\infty} \phi_i(\vec{r}) \chi_i(\vec{R}).
\label{weinberg2}
\end{equation}
When  this expansion is introduced into the three-body equation Eq.(\ref{3b-eq}), one obtains
a set of coupled channel equations:
\begin{eqnarray}
& & [E+i\epsilon -  K_\mathbf{R}  - \bar U_{ii}(\vec R)] \chi_i(\vec R) = \nonumber \\
& & \imath \epsilon \delta_{i1} N_d \exp(\imath \vec K_d .\vec R) +
\sum_{j \ne i} \bar U_{ij} (\vec R) \chi_{j}(\vec R). \,
\label{jt1-eq}
\end{eqnarray}
The coupling potentials are defined by $\bar U_{ij}(\vec R) = U_{ij} + \beta_{ij} (\alpha_j -1)$
and $U_{ij}(\vec R) = - \langle \phi_i | V_{np} ( U_{nA} + U_{pA} | \phi_j \rangle$ where
$\beta_{ij} = \langle \phi_i | V^2_{np}  | \phi_j \rangle $ and $\alpha_j$ are the eigenvalues of the
Wienberg equation Eq.(\ref{weinberg1}).
The normalization coefficient appearing on the r.h.s of Eq.(\ref{jt1-eq}) is
$N_d=-\langle \phi_1 | V_{np} | \phi_d \rangle$.

These coupled channel equations can be solved exactly as done in \cite{laid} but reduce to a much
simpler form if only the first term of the expansion Eq.(\ref{weinberg2}) is necessary.
In that case, $\alpha_1=1$ and we can arrive at the following optical model type equation:
\begin{equation}
(E + \epsilon_d + i \varepsilon - T_R - U_{11}(R)) \chi^{JT}_1(\vec{R})= i \varepsilon N_d  \, \exp(\imath \vec K_d .\vec R) ,
\label{jt2-eq}
\end{equation}
where now the  potential is still related to the sum of the proton and neutron
potentials as in Johnson and Soper, but involves a more complex folding procedure:
\begin{equation}
U_{11}(R)= - \langle \phi_1(\vec{r}) | V_{np} (U_{nA}+U_{pA}) | \phi_1(\vec{r}) \rangle \, .
\label{jtpot-eq}
\end{equation}
Apart from the normalization, $\phi_1$ is the ground state wavefunction
of the deuteron $\phi_d$. Then the potential $U_{11}(R)$ can also be written in terms of $\phi_d$:
\begin{equation}
U_{11}(R)\equiv U_{JT}(R) =  \frac{\langle \phi_d(\vec{r}) | V_{np} (U_{nA}+U_{pA}) | \phi_d(\vec{r}) \rangle}{\langle \phi_d(\vec{r}) | V_{np} | \phi_d(\vec{r}) \rangle} \, .
\label{jtpot2-eq}
\end{equation}

In this case the transfer amplitude is defined through the 6-dimensional integral:
\begin{eqnarray}
T &=& <\phi_{nA}\chi^{(-)}_{pB}| V_{np} | \phi_1 \chi^{JT}_1(\vec{R})>\nonumber \\
 &=&<\phi_{nA}\chi^{(-)}_{pB}| V_{np} | \phi_d (\chi^{JT}_1(\vec{R})/N_d)>.
\label{tmatrix-fr}
\end{eqnarray}
where we have used $\mid \phi_d\,> =N_d \mid \phi_1>$.

We see from Eq.(\ref{jt2-eq}) that $(\chi^{JT}_1(\vec{R})/N_d)$ is a distorted wave generated by the potential $U_{11}$ and normalized to an incident wave of unit amplitude.

In the following section we will compare the cross sections obtained with
Eq.(\ref{tmatrix-zr}) and Eq.(\ref{tmatrix-fr}). We also disentangle the separate
effects of finite range in the potential $U_{11}$, looking
specifically at the potentials from Eq.(\ref{jspot-eq}) and (\ref{jtpot-eq}),
and that of finite range in the evaluation of the transfer amplitude.


\section{Results}

We perform a systematic study of finite range effects on 26 (d,p) reactions.
In all our calculations we take the Reid interaction
for the deuteron \cite{rsc} and the Chapel Hill global parameterization for the
nucleon optical potentials \cite{ch89}. In calculating the potential of Eq.\ref{jtpot2-eq}
we neglect the d-wave part of the $\phi_d$. For all cases here presented,
the final bound single particle state is obtained using a potential with
standard radius and diffuseness $r=1.2$ fm and $a=0.65$ fm and adjusting
the depth to the known neutron separation  energy of the corresponding final state.
In Subsection(IIIA) we show our results for the  potential  and
then present the results for the (d,p) cross sections in Subsection(IIIB).

\subsection{Finite range effects in the potentials}

Johnson and Tandy potentials Eq.(\ref{jtpot2-eq}) are computed, using a subroutine contained in the code TWOFNR \cite{twofnr} for performing the $r$ integrations needed,
 and compared with the Johnson and Soper adiabatic potentials Eq.(\ref{jspot-eq}) for 26 cases. In order to simplify the
comparison we fit the real part of the resulting $U_{JT}$ to a volume
Woods Saxon shape and the corresponding imaginary to a surface Woods Saxon form.
For all cases studied
we find that the most important difference between the interactions $U_{JT}$ and $U_{JS}$
is a constant increase in diffuseness. There is also a slight systematic decrease in radius.
Differences in the depths of the real and imaginary parts are more subtle and
vary case to case. In \cite{wales}, an approximate method to estimate
finite range corrections to the adiabatic potential was developed. In that method,
the radius is fixed but an increase in diffuseness is predicted with a decrease in the depth
of the potential.
In table \ref{table-potential} we show the percentage difference of our numerically calculated
$U_{JT}$  and the Wales and Johnson approximate prescription $U_{WJ}$ \cite{wales}, relative to
the zero-range Johnson and Soper potential $U_{JS}$, for three reference cases.
Percentage differences are calculated at the radius for which the potential is maximum.

The main feature of $U_{JT}$
compared to $U_{JS}$ is captured by the Wales and Johnson prescription, namely the increase
in the diffuseness in both the real and imaginary part of the interaction. However the Wales and Johnson results differ quantitatively from ours.

\begin{table}[htdp]
\caption{Finite range effects on the Johnson-Tandy distorting potential, Eq.(\ref{jtpot2-eq}): in the 3rd column the Wales and Johnson potential is compared with the zero range potential $U_{JS}$ and in the 4th column the Johnson and Tandy $U_{JT}$
is compared with $U_{JS}$. We compare the diffuseness of the real and imaginary parts, $a_R$ and $a_I$,
as well as the depths of the real and imaginary parts $V$ and $W_s$, and the corresponding
radii $r_R$ and $r_I$. }
\label{table-potential}
\begin{center}
\begin{tabular}{|c|c|c c|}					
\hline
target 	&   parameter  &  $U_{WJ}$    	      & $U_{TJ}$   	\\
\hline
all		&$a_R$  &  +4\%                & +7\%              \\
		&$a_I$  &  +3\%                & +8-9\%               \\
\hline
$^{12}$C 	&$V$    &  -5.6\%              & -1.98\%              \\
	 	&$r_R$  &  0\%                 & -1.25\%              \\
		&$W_s$  &  -4.6\%              & -4.52\%               \\
		&$r_I$  &  0\%                 & +0.72\%               \\
\hline
$^{48}$Ca  	&$V$    &  -2.1\%              & -0.04\%              \\
	 	&$r_R$  &  0\%                 & -0.93\%              \\
	  	&$W_s$  &  -3.7\%              & +1.6\%               \\
	 	&$r_I$  &  0\%                 & -0.97\%              \\
\hline
$^{208}$Pb 	&$V$    & -0.7\%              & +0.06\%              \\
	 	&$r_R$  &  0\%                 & -0.35\%              \\
	 	&$W_s$  & -3.3\%              & +1.2\%               \\
	 	&$r_I$  &  0\%                 & -0.35\%              \\ \hline
\end{tabular}
\end{center}
\label{default}
\end{table}

\subsection{Finite range effects in the transfer cross sections}

\begin{table}
\caption{\label{fr-effects} Percentage differences of finite range effects in (d,p) reactions
relative to the zero-range Johnson and Soper calculation (* denotes cases for which no data
is available). The target nucleus, the deuteron incident energy in the laboratory (in MeV) and the angle (in degrees) at which the percentage differences are evaluated are given in the 1st, 2nd and 3rd column respectively. }
\vspace{0.25cm}
{\centering
\begin{tabular}{|l|r|r|r|r|r|r|}					\hline
Target 		&  $E_d$ &$\theta$ & $\Delta$(LEA)	& $\Delta$(FR-JS)& $\Delta$(FR-JT) &$\Delta$(JT-JS)\\  \hline
$^{12}$C 	& 4  & 25  &+5.6\%    	&+5.5\%   	&+4.5\% 	& -1.0\%     \\
$^{12}$C 	& 12 & 13  &+2.6\%    	&+2.9\%   	&-1.5\% 	& -4.3\%\\
$^{12}$C 	& 19.6& 10 &+11\%     	&+13\%  	&+7.7\% 	& -4.2\%\\
$^{12}$C 	& 56 & 6   &-37\%    	&-27\%    	&-36\%  	&-12\%\\
\hline
$^{48}$Ca 	& 2 & 180  &+6.5\%    	&+6.3\%   	&+2.6\% 	&-3.5\%	\\
$^{48}$Ca 	& 13 & 12  &+4.9\%    	&+3.8\%   	&-2.8\% 	&-6.2\%\\
$^{48}$Ca 	& 19 & 8   &+5.0\%     	&+4.0\%     	&-0.30\% 	&-4.1\%\\
$^{48}$Ca* 	& 30 & 4   & 	    	&+7.3\%   	&+4.8\%  	&-2.3\%\\
$^{48}$Ca* 	& 40 & 0   &	    	&-5.4\%   	&-5.9\%  	&-10\%\\
$^{48}$Ca* 	& 50 & 0   &	    	&-1.9\%   	&-19\%  	&-18\%\\
$^{48}$Ca 	& 56 & 0   &-5.2\%    	&-6.5\%   	&-24\%  	&-18.6\%\\
\hline
$^{69}$Ga 	& 12 & 14  &+4.3\%      &+4.7\%   	&-1.1\%		&-5.49\% \\
\hline
$^{86}$Kr 	& 11 & 25  &+4.8\%    	&+5.5\%   	&-0.40\%	&-5.63\% \\
\hline
$^{90}$Zr 	& 2.7& 138 &+6.2\%	&+7.3\%		&+5.5\%  	&-1.7\%\\
$^{90}$Zr 	& 11 & 26  &+5.4\%    	&+5.0\%      	&-0.90\%      	&-5.6\%\\
\hline
$^{124}$Sn	& 5.6 & 175&+6.1\% 	&+11\%		&+7.5\% 	&-2.8\%\\
$^{124}$Sn	& 33.3& 0  &+2.9\%    	&+4.6\%    	&0\%        	&-4.4\% \\
$^{124}$Sn*	& 40 & 12  &	   	&-1.1\%  	&-2.4\%   	&-1.4\%\\
$^{124}$Sn*	& 50 & 11  &	   	&-3.9\%  	&-4.3\%   	&-0.44\%\\
$^{124}$Sn*	& 60 & 9   &	   	&-11\%  	&-30\%   	&-21\%\\
$^{124}$Sn*	& 70 & 0   &+5.1\%   	&-29\%  	&-44\%   	&-21\%\\
\hline
$^{208}$Pb	& 8 &  180 &+6.1\%    	&+7.2\%   	&+6.1\% 	&-0.96\%\\
$^{208}$Pb	& 12 &  98 &+5.7\%    	&+8.8\%   	&+2.2\% 	&-6.1\%\\
$^{208}$Pb*	& 20 & 30  &	    	&+4.5\%   	&-2.3\% 	&-6.6\%\\
$^{208}$Pb*	& 40 & 9   &	    	&+1.4\%   	&-6.9\% 	&-8.1\%\\
$^{208}$Pb*	& 60 & 0   &	    	&+0.14\%   	&-8.8\% 	&-9.0\%\\
$^{208}$Pb*	& 80 & 0   &	   	&-62\%   	&-86\% 		&-63\%\\
\hline
\end{tabular}}
\end{table}
Once the adiabatic potentials are defined, cross sections can be obtained.
The matrix element in Eq.(\ref{tmatrix-zr}) was evaluated using the adiabatic wavefunction
distorted by $U_{JS}$ and the zero-range constant $D_0$ was obtained from
the Reid n-p interaction \cite{rsc} for consistency.
The finite range calculation Eq.(\ref{tmatrix-fr}) uses  $U_{JT}$ for the  adiabatic wavefunction.

To assess the relative importance of the finite range effect in the adiabatic
potential and that on the evaluation of the transfer amplitude, we also perform
a finite range calculation using $U_{JS}$. Finally, given that the LEA method
\cite{lea} has been widely used in the past, we also perform a calculation
where Eq.(\ref{tmatrix-zr}) is evaluated using the Johnson and Soper adiabatic potential
$U_{JS}$ but making the local energy correction \cite{lea}.
All calculations are performed using the code {\sc fresco} \cite{fresco}.

In table \ref{fr-effects} we quantify these effects for all the reactions studied.
All percentage differences are calculated at the first peak of the angular distribution
(with the exception of the sub-Coulomb examples for which percentage differences
are calculated at backward angles) and are relative to the Johnson and Soper approach Eq.(\ref{tmatrix-zr}):
\begin{itemize}
\item $\Delta$(LEA) shows the effect of finite range in the evaluation of the
T-matrix when the LEA is used in conjunction with the Johnson and Soper model;
\item $\Delta$(FR-JS) shows the effect of finite range in the evaluation of the
T-matrix when fixing the  adiabatic potential to $U_{JS}$;
\item $\Delta$(FR-JT) shows the full finite range effects when finite range
is taken into account properly both in the evaluation of the
T-matrix and the  adiabatic potential ($U_{JT}$);
\item $\Delta$(JS-JT) is the percentage difference between finite range calculations
using $U_{JT}$ and $U_{JS}$ and therefore shows the effect of including finite range
in the adiabatic potential.
\end{itemize}
In the table \ref{fr-effects} we indicate the angle at which the percentage differences
of cross sections were calculated.

\begin{figure}[t!]
{\centering \resizebox*{0.42\textwidth}{!}{\includegraphics{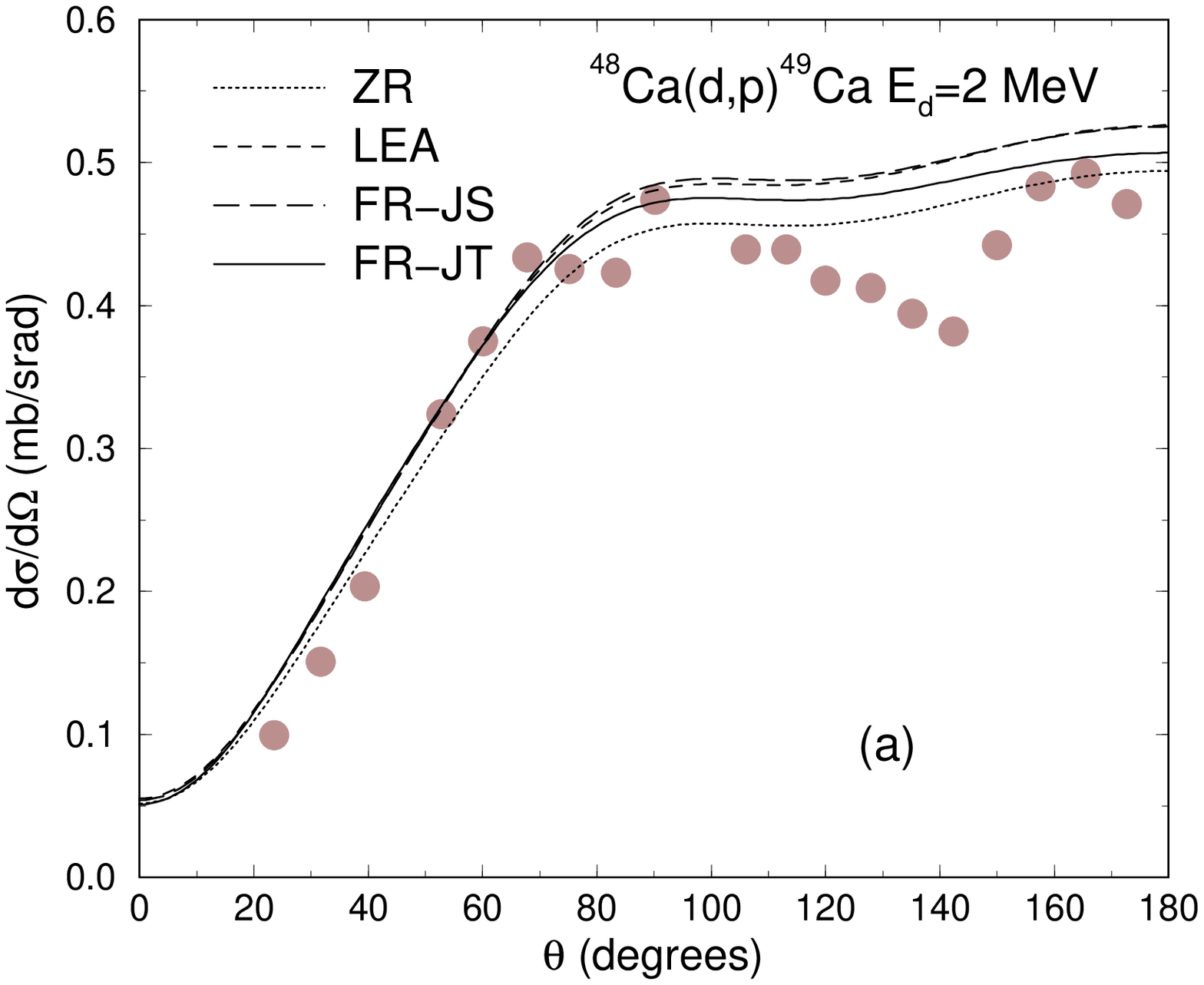}}} \\
{\centering \resizebox*{0.42\textwidth}{!}{\includegraphics{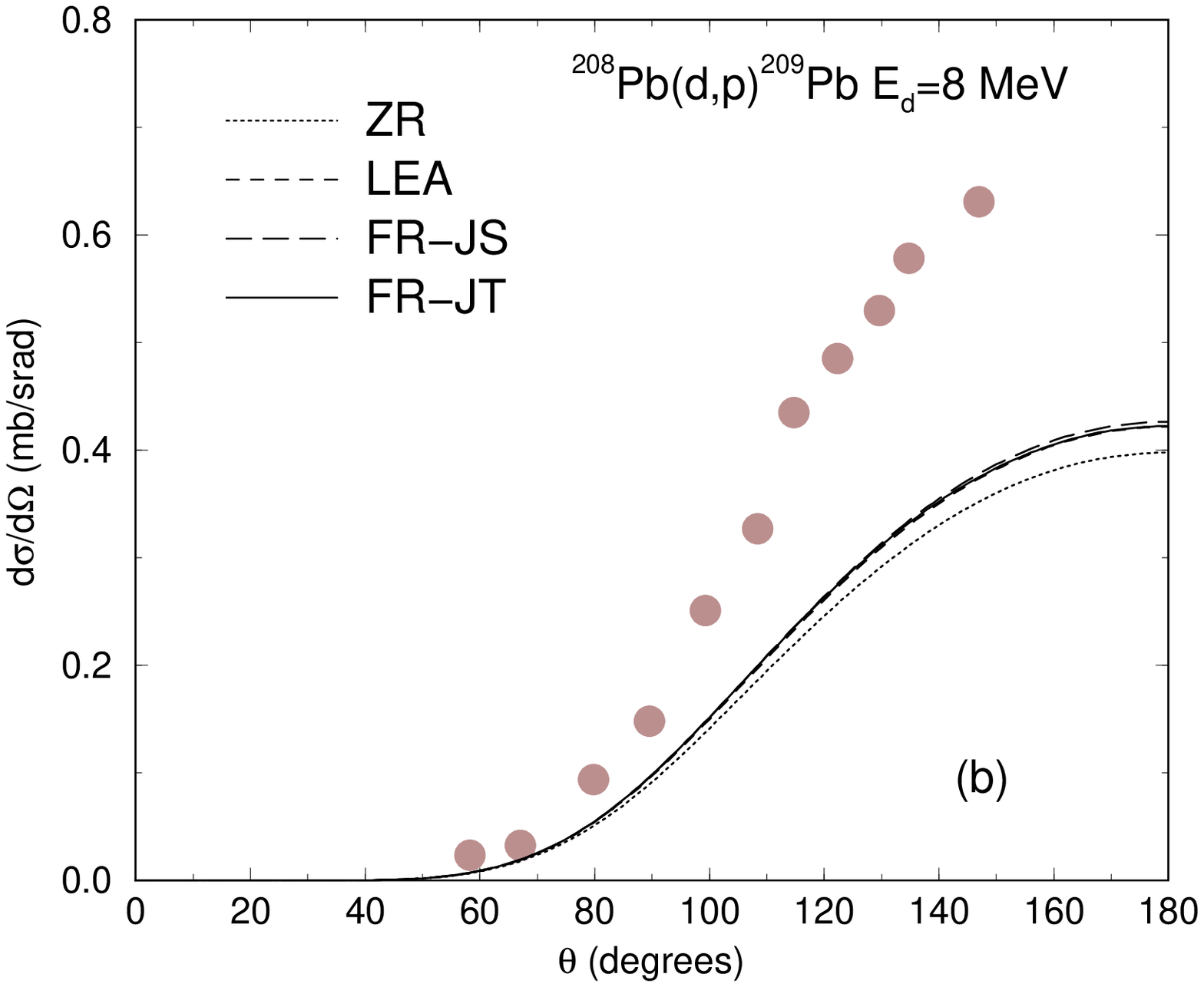}}}
\caption{\label{dsde-sub} Angular distributions for (d,p)
at sub-Coulomb energies: (a) $^{48}$Ca(d,p)$^{49}$Ca(g.s.) $E_d = 2$ MeV (data from \cite{ca48-2}) and (b) $^{208}$Pb(d,p)$^{209}$Pb(g.s.) $E_d = 8$ MeV (data from \cite{pb208-8}).
Comparison of zero-range Johnson and Soper model (dotted),
the LEA Johnson and Soper model (dashed), a finite range
calculation of the transfer amplitude using the Johnson and Soper adiabatic
wave (long-dashed) and the full finite range results (solid line). }
\end{figure}

\begin{figure}[!t]
{\centering \resizebox*{0.42\textwidth}{!}{\includegraphics{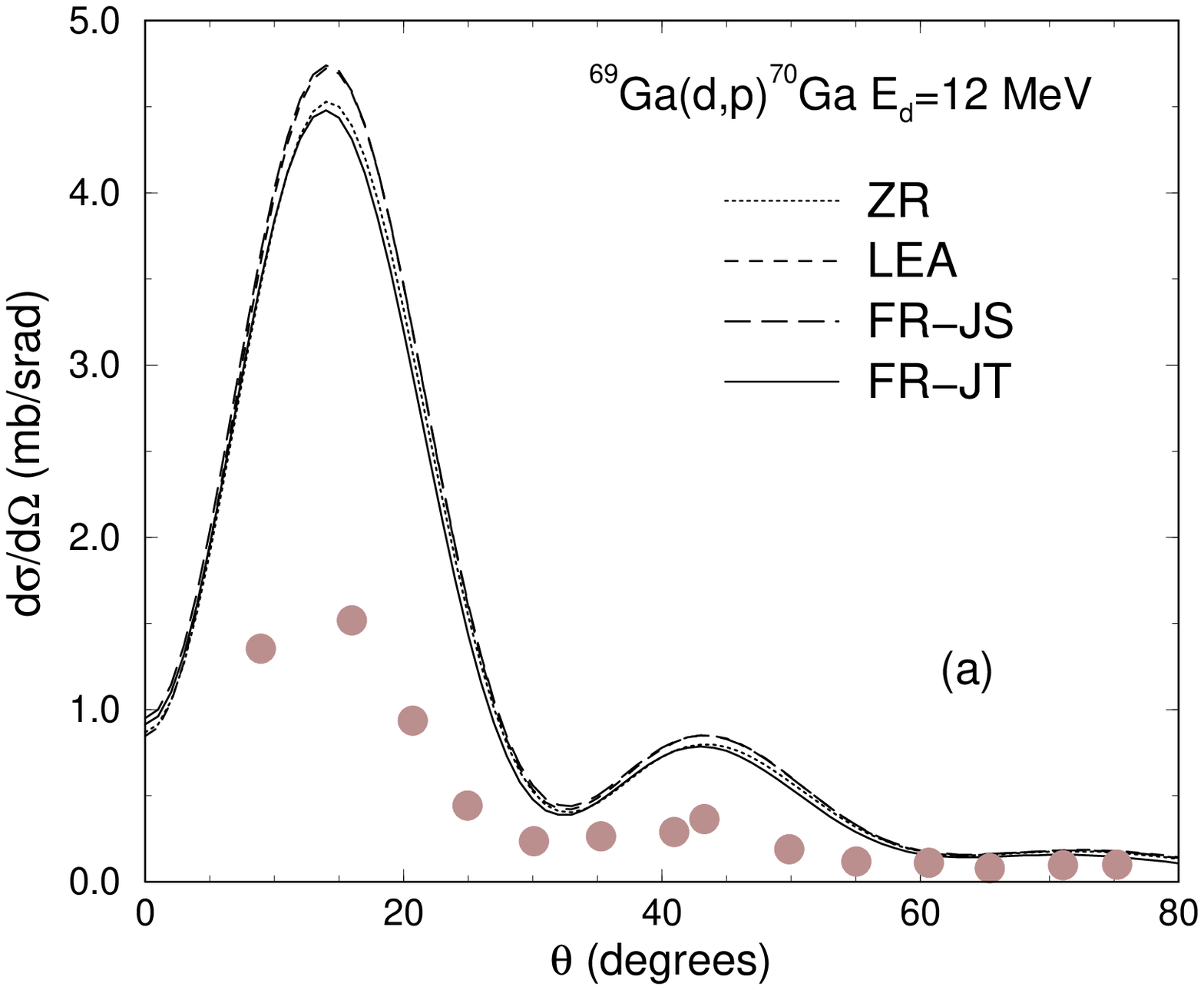}}} \\
{\centering \resizebox*{0.42\textwidth}{!}{\includegraphics{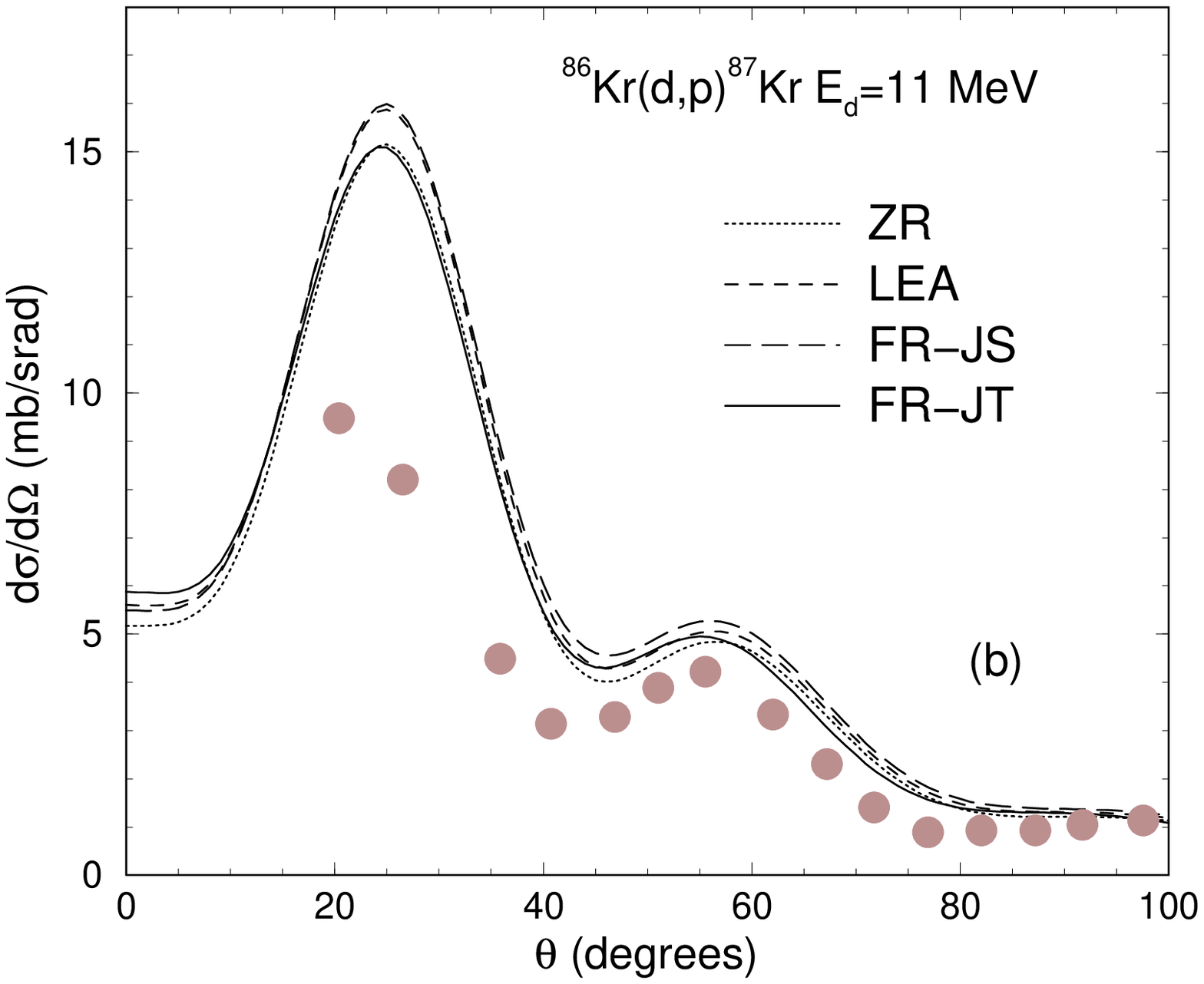}}} \\
{\centering \resizebox*{0.42\textwidth}{!}{\includegraphics{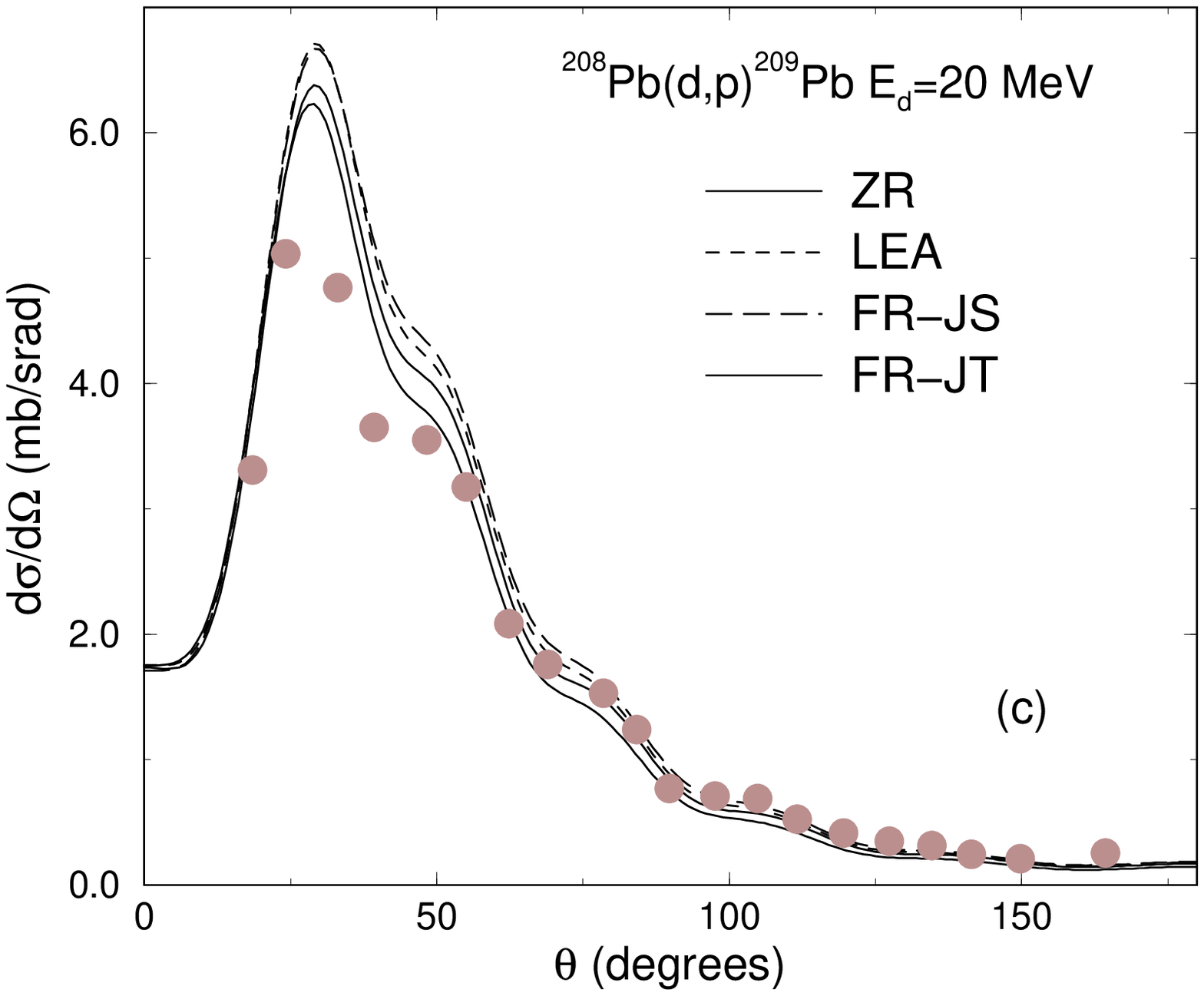}}}
\caption{\label{dsde-inter} Angular distributions for (d,p)
at energies slightly above the Coulomb barrier:
(a) $^{69}$Ga(d,p)$^{70}$Ga(g.s.) $E_d = 12$ MeV (data from \cite{ga69}),
(b) $^{86}$Kr(d,p)$^{87}$Kr(g.s.) $E_d = 11$ MeV (data from \cite{kr86}) and (c)  $^{208}$Pb(d,p)$^{209}$Pb(g.s.) $E_d = 20$ MeV (data from \cite{pb208-20}).
Comparison of zero-range plus Johnson and Soper method (dotted),
the LEA plus Johnson and Soper method (dashed), a finite range
calculation of the transfer amplitude using the Johnson and Soper adiabatic
potential (long-dashed) with the full finite range results (solid line).
}
\end{figure}

\begin{figure}[t!]
{\centering \resizebox*{0.42\textwidth}{!}{\includegraphics{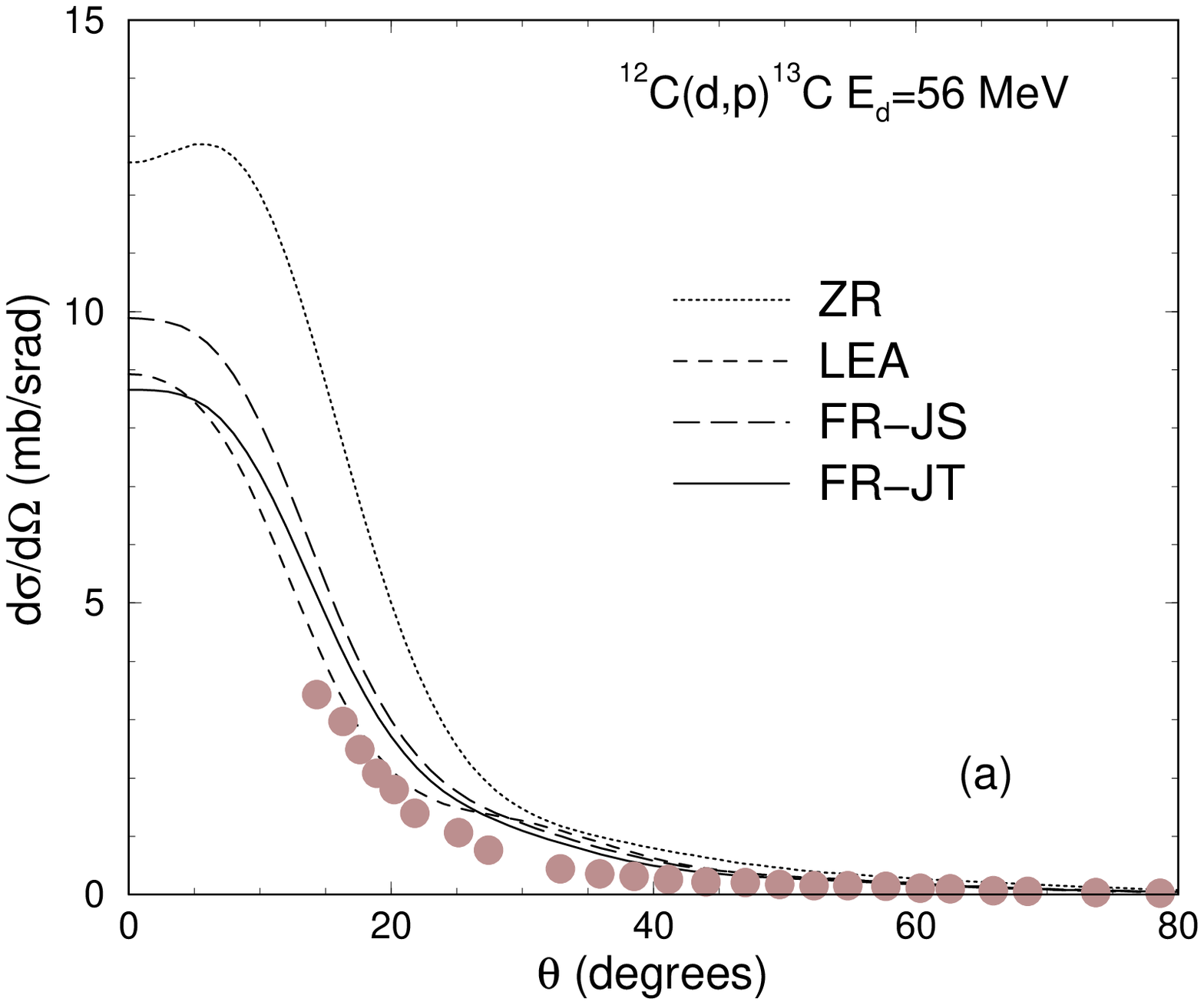}}} \\
{\centering \resizebox*{0.42\textwidth}{!}{\includegraphics{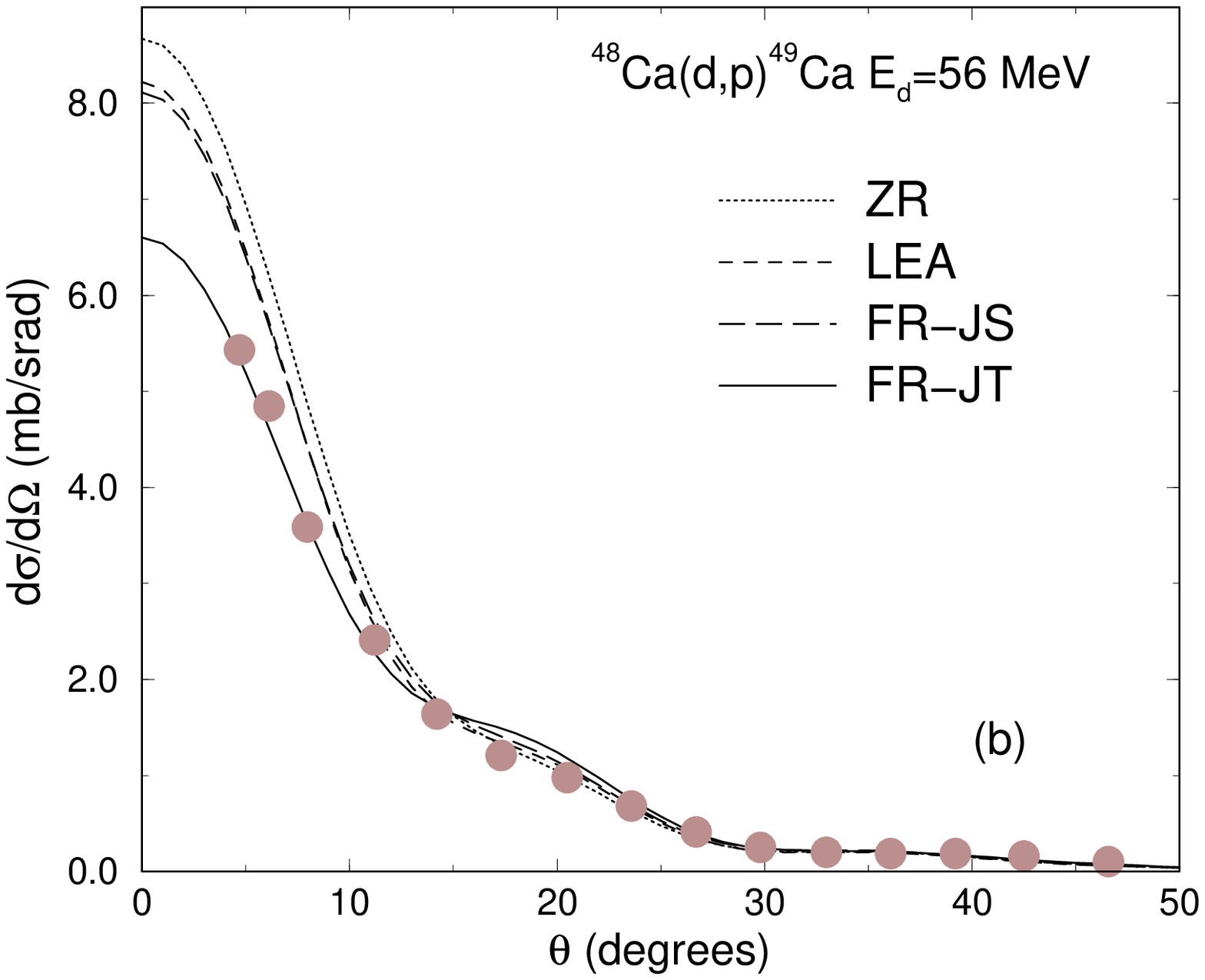}}} \\
{\centering \resizebox*{0.42\textwidth}{!}{\includegraphics{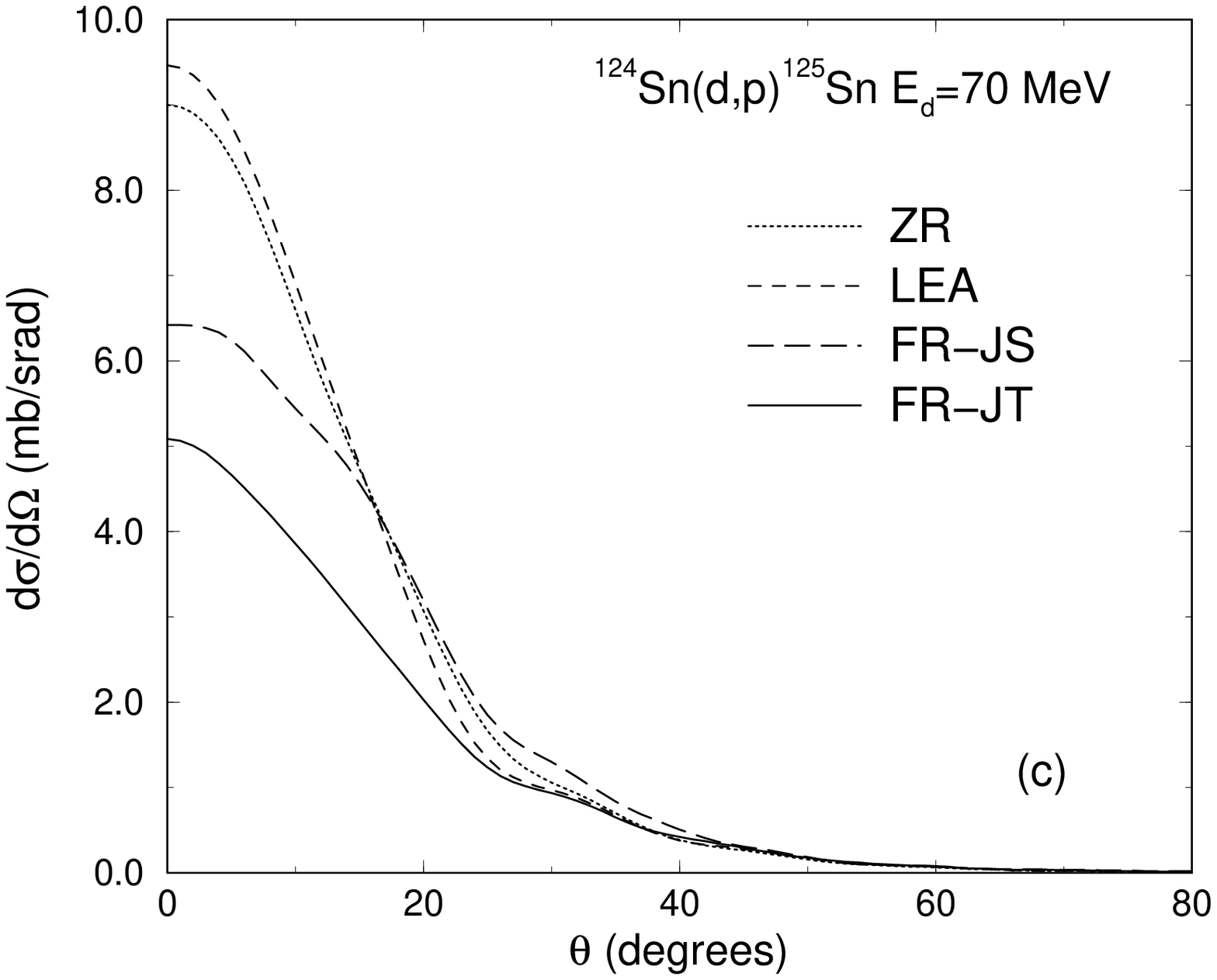}}}
\caption{\label{dsde-high} Angular distributions for (d,p)
at high energies: (a) $^{12}$C(d,p)$^{13}$C(g.s.) $E_d = 56$ MeV (data from \cite{c12-56}), (b) $^{48}$Ca(d,p)$^{49}$Ca(g.s.) $E_d = 56$ MeV (data from \cite{ca48-56}) and (c)  $^{124}$Sn(d,p)$^{125}$Sn(g.s.) $E_d = 70$ MeV.
Comparison of zero-range plus Johnson and Soper method (dotted),
the LEA plus Johnson and Soper method (dashed), a finite range
calculation of the transfer amplitude using the Johnson and Soper adiabatic
potential (long-dashed) with the full finite range results (solid line). }
\end{figure}

In addition to the full table, we also select some angular distributions that illustrate
the various trends we observed, presented in Figs.\ref{dsde-sub}-\ref{dsde-high}. Each plot contains four lines: a dotted line
corresponding to the zero-range Johnson and Soper calculation, a dashed line corresponding to
the local energy correction to the $T$-matrix calculation with the Johnson and Soper potential, the long-dashed line corresponding
to a full finite range $T$-matrix calculation where $U_{JS}$ is used, and the full line is a finite range
$T$-matrix calculation with the finite range adiabatic potential  $U_{JT}$. Data is also presented whenever available,
but only as an indication that the ingredients of our model are realistic and therefore the magnitude
of the finite range effects reliable. It is not the purpose of this work to extract
spectroscopic information for these systems.

We first look at sub-Coulomb transfer reactions, which are usually rather insensitive
to the nuclear optical potential. In Fig.\ref{dsde-sub}(a) we show $^{48}$Ca(d,p)$^{49}$Ca at $E_{lab}=2$ MeV and in Fig.\ref{dsde-sub}(b) $^{208}$Pb(d,p)$^{209}$Pb
at $E_{lab}=8$ MeV. In both cases, the effects of finite range are only a few percent
(3\% in $^{48}$Ca and 6\% in $^{208}$Pb), and these are mostly due to the approximation
in the evaluation of the T-matrix and not from the  adiabatic potential.
In this case, LEA is able to capture most of the finite range effects.
In the sub-Coulomb energy regime, for all cases studied, errors
in using the Johnson and Soper potential with the local energy approximation
are below 5\%.

Most of the available (d,p) data on stable systems was taken at energies above the Coulomb barrier
for 10-20  MeV deuterons. In Fig. \ref{dsde-inter}(a) we show results for
$^{69}$Ga(d,p)$^{70}$Ga at $E_{lab}=12$ MeV, in Fig. \ref{dsde-inter}(b)
$^{86}$Kr(d,p)$^{87}$Kr at $E_{lab}=11$ MeV and in Fig. \ref{dsde-inter}(c) $^{208}$Pb(d,p)$^{209}$Pb at $E_{lab}=20$ MeV.
The overall effect of finite range in all three cases is very small
($-1\%$ for the $^{69}$Ga, $0.4\%$ for the $^{86}$Kr and 6\% in $^{208}$Pb), although it results
from the cancellation of the two separate effects, the finite range in the deuteron potential,
which reduces the cross section and the finite range effect in the evaluation of the T-matrix
which increases the cross section. No simple addition rule for these two effects was found.
Here, the local energy approximation begins to show larger deviations from the
full finite range calculation.

Finally, we also consider reactions at higher energies (50-80 MeV deuteron energy).
Only two data sets are available, namely for $^{12}$C and $^{48}$Ca but we
also include a study for $^{124}$Sn and $^{208}$Pb to ensure that our results are not biased
by lower mass systems. All cases studied at these energies reveal that finite range
effects are large and reduce the cross section. In Fig. \ref{dsde-high}(a) we show
the angular distributions for $^{12}$C(d,p)$^{13}$C at $E_{lab}=56$ MeV, in Fig. \ref{dsde-high}(b) those for $^{48}$Ca(d,p)$^{49}$Ca
at $E_{lab}=56$ MeV and in Fig. \ref{dsde-high}(c) those for  $^{124}$Sn(d,p)$^{124}$Sn
at $E_{lab}=70$ MeV. The overall effect of finite range at the peak of the distribution
is $-36\%$ for $^{12}$C, $-24\%$ for $^{48}$Ca and $-43.5$\% for $^{124}$Sn.
It is the finite range in the adiabatic potential that is the dominant cause for
these large changes although the finite-range effect in the evaluation of the T-matrix
is still important and should not be neglected. In addition,
for these higher energies, we find that the LEA method breaks down and
for the heavier systems this approximation can in fact provide a correction in the opposite
direction to the full finite range calculation.

Because it is the adiabatic scattering wavefunction that is mostly responsible
for the large differences, we investigated the radial behavior of the
scattering wavefunctions using either $U_{JT}$ or $U_{JS}$ for the partial waves which
contribute the most to the transfer cross section. We specifically looked at the properties
of the integrand of Eq.(\ref{tmatrix-fr}) in the zero-range approximation where
it has a simpler form. We found that the percentage difference
comes from subtle cancellations and cannot be
well illustrated in the partial wave expansion. Intuitively one might argue that
since the energies are larger, the dominant contribution to the transfer cross
section comes from smaller impact parameters and thus
sensitivity to the range of $V_{np}$ should be larger.

To ensure that our results are general, in particular that they will still be applicable
to reactions in which the final bound state has a large spatial extension, we performed additional calculations
for a fictitious $^{48}$Ca(d,p)$^{49}$Ca setting the valence neutron angular
momentum in the final bound state to $\ell=0$ state, and varying the binding energy
$S_n=0.1-6$ MeV. The overall findings did not change: regardless of the loosely
bound nature of the final nucleus,
or the angular momentum in the final bound state, the effects of finite
range in the transfer cross section are modest for low energies and
become very important for the higher energies.

\section{Discussion}

\begin{figure}[t!]
{\centering \resizebox*{0.42\textwidth}{!}{\includegraphics{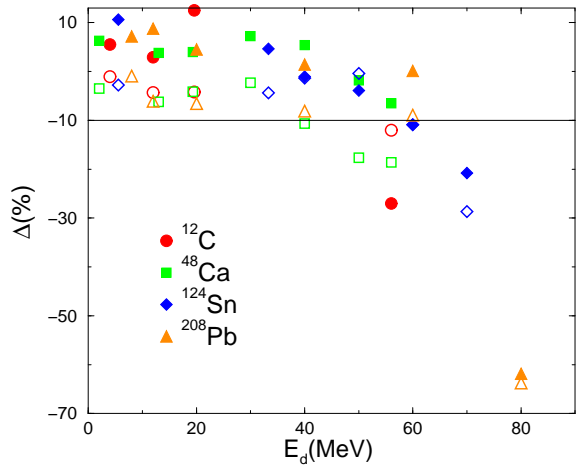}}} \\
\caption{\label{systematic-fig} Systematic finite range effect as a function
of beam energy: open symbols give the effect in the incident channel distorted wave,
and the filled symbols are the effect in the evaluation of the matrix elements.}
\end{figure}

The overall features obtained in this work can be best summarized in
Fig.\ref{systematic-fig} where the two separate effects of finite range
are plotted as a function of beam energy for (d,p) reaction on four different targets:
solid symbols provide the percentage effect of including finite range in the evaluation
of the matrix element relative to a zero-range $T$-matrix calculation with a Johnson and Soper potential in both cases, and the open symbols
correspond to the effect of including finite range effects in the distorting potential in the incident channel wave.
The figure summarizes the results already given in Table \ref{fr-effects}.
From this figure we can see that at low energy, finite range results differ  by less than 10\%
 from zero range  matrix element with a Johnson and Soper adiabatic potential.
However, as the  incoming deuteron energy increases, finite range effects become very important and
can dominate the result.
The energy at which the transition occurs depends non linearly
on the charge and the mass of the system. For practical purposes we find
the transition to be around 20 MeV/u for
lighter systems and 30 MeV/u for the heavy systems.

To facilitate the practical analysis of  experiments, we searched for a global correction factor,
a factor that would estimate the finite range effect as a function of target charge, mass
and beam energy. Although for a given target one could always find a function $F(E_d)$
representing the finite-range correction,
no consistent dependence in mass and charge was found
for the parameters of the various fits.

It is important to remember that the exact inclusion of deuteron finite range effects
requires the solution of a couple channel equations Eq.(\ref{jt1-eq}) and here we
truncated the Weinberg expansion to the first term to simplify the problem.
The full equations were solved in \cite{laid} for $^{66}$Zn.
It would be interesting to extend this study to better determine the
range of validity of the truncation here used.

\section{Conclusion}

We perform a systematic study of the effects of deuteron finite range in (d,p) reactions,
within a formalism that includes the coherent effects of deuteron breakup through an adiabatic potential in the incident channel. We use the adiabatic formalism developed by Johnson and Soper \cite{soper} in zero range
and compare with the finite range generalization of Tandy and Johnson \cite{tandy}.
We analyze separately the effects of finite-range in the adiabatic distorting potential and finite range in the evaluation of the transfer matrix element.
We also test the local energy approximation which is widely used as an estimate of finite range  corrections to zero range transfer cross sections. We performed (d,p)
calculations to determine angular distributions
for a wide range of beam energies as well as a variety of targets, from A=12 to A=208.

For sub-Coulomb reactions,
the percentage difference between the finite range and the zero range cross sections
at the peak of the angular distribution relative to the zero range Johnson and Soper
prediction is within 10\% for all cases studied, and the local energy approximation
provides an estimate within a few percent of the full finite range calculation.
However, as the beam energy increases,
finite range effects become more important. For intermediate energies ($E<20$ MeV/u for $A<50$
and $E<30$ MeV/u for heavier nuclei),
including the finite range of the $n-p$ interaction in the adiabatic scattered wavefunction reduces
the cross section while  including finite range in the evaluation of the transfer
amplitude increases the cross section. Both effects are significant,  although
strong cancellations may occur. At higher energies,
both finite range effects have the same sign, reducing the transfer
cross section. In this case we find the total effect of finite range to be very
important. Our results also suggest that at these higher energies,
the local energy approximation is no longer adequate.

This work was partially supported by the National Science Foundation
grant PHY-0555893 and the Department of Energy under contract DE-FG52-08NA28552. 
RCJ is also partially supported by the United Kingdom Science and Technology Facilities 
Council under Grant No. ST/F012012.


\end{document}